\documentclass[amsfonts,amssymb,amsmath,prb,aps,reprint,notitlepage,superscriptaddress,longbibliography]{revtex4-2}
\usepackage[T1]{fontenc}
\usepackage{times}
\usepackage{xcolor}
\usepackage[colorlinks,allcolors=blue]{hyperref}
\usepackage{graphicx}
\usepackage{bm}
\usepackage{mathrsfs}
\usepackage{dcolumn}
\usepackage{float}
\usepackage{dsfont}
\usepackage{amsmath}
\usepackage{amsfonts}
\usepackage{amssymb}
\usepackage{braket}
\usepackage{stmaryrd}

\newcommand{\norm}[1]{\left\lVert#1\right\rVert}

\renewcommand{\vec}[1]{{\ensuremath{\bm{#1}}}}

\newcommand{\muB}{{\ensuremath{\mu_{\mathrm{B}}}}}
\begin{document}

\title{Path integral spin dynamics with exchange and external field}

\author{Thomas Nussle}
\email{t.s.nussle@leeds.ac.uk}
\affiliation{School of Physics and Astronomy, University of Leeds, United Kingdom}
\author{Stam Nicolis}
\email{stamatios.nicolis@univ-tours.fr}
\affiliation{Institut Denis Poisson, Universit\'{e} de Tours, Universit\'{e} d'Orl\'{e}ans, CNRS (UMR7013), Parc de Grandmont, F-37200, Tours, FRANCE}
\author{Iason Sofos}
\affiliation{School of Physics and Astronomy, University of Leeds, United Kingdom}
\author{Joseph Barker}
\email{j.barker@leeds.ac.uk}
\affiliation{School of Physics and Astronomy, University of Leeds, United Kingdom}

\begin{abstract}
In this work, we propose a path integral-inspired formalism for computing the quantum thermal expectation values of spin systems, when subject to magnetic fields that can be time-dependent and can accommodate the presence of Heisenberg exchange interactions between the spins. This is done by deriving an effective magnetic field from the quantum partition function of the system to use in classical atomistic spin dynamics simulations and generalises the formalism presented in our previous work [Phys. Rev. Research 5, 043075 (2023)]. In special cases where the effective field can be computed exactly we compare our results with exact/numerical diagonalisation methods for both ferromagnetic and antiferromagnetic coupling. We show that our method works well across a large temperature range and can reproduce quantum expectation values for antiferromagnetic coupling which is usually not possible with classical models.  
\end{abstract}

\maketitle

%%%%%%%%%%%%%%%%%%%%%%%%%%%%%%%%%%%%%%%%%%%%%%%%%%%%%%%%%%%%%%%%%%%%%%%%%%%%%%%%
\section*{Introduction\label{introduction}}
%%%%%%%%%%%%%%%%%%%%%%%%%%%%%%%%%%%%%%%%%%%%%%%%%%%%%%%%%%%%%%%%%%%%%%%%%%%%%%%%
  Path integrals have been a rich source of inspiration ever since Feynman proposed them as an alternative approach to solving and interpreting quantum mechanical problems \cite{feynmanSpaceTimeApproachNonRelativistic1948}. Even in recent years, numerical methods based on Feynman's path integral formalism have continued to be developed and have proven to be particularly efficient, as, for example, in the field of path integral molecular dynamics \cite{marxPathintegralMonteCarlo1993, marxInitioPathIntegral1996, ceriottiEfficientStochasticThermostatting2010}. Beyond the standard position and conjugate momentum description of quantum mechanics, a path integral for spin systems was also written long ago and has already been thoroughly investigated \cite{schulmanPathIntegralSpin1968, klauderPathIntegralsStationaryphase1979, cabraPathIntegralRepresentation1997a}. These efforts show how constraints can be taken into account in the path integral formalism. In previous work, we established a framework, inspired by path integral molecular dynamics methods, specifically designed for spin systems. We studied the case of a single spin, coupled to a constant external field, through the Zeeman interaction \cite{nussleNumericalSimulationsSpin2023}. Within this framework, we derived an effective classical Hamiltonian that captured the quantum fluctuations--as well as the thermal fluctuations, upon coupling it to a thermal bath. Using a systematic double expansion of the partition function in powers of $1/\mathrm{spin}$ and $\hbar\times\mathrm{spin}$, we computed approximate spin expectation values through numerical stochastic atomistic spin dynamics simulations, which we compared to exact quantum thermal expectation values. In this work, we extend the formalism in two ways: (i) introducing interactions between spins on different sites through an isotropic Heisenberg exchange Hamiltonian and (ii) allowing the Zeeman interaction of a field in a general direction, to couple the spin system to a potentially time dependent external magnetic field. Both of these extensions introduce additional complexities arising from the non-commutativity of the spin operators due to the curvature of the group manifold, on the one hand, and quantum effects, on the other hand. This intricate interplay between classical--geometric--and quantum effects is what we aim to incorporate in our effective Hamiltonian for the classical spin system, now upon taking into account the complexity of the exchange interaction.

Our work has its natural place in the context of semiclassical approaches to the treatment of quantum problems through the mapping of quantum problems to stochastic classical approximations, which dates back at least as far as 1968 \cite{laxQuantumNoiseXI1968, gilmoreClassicalquantumCorrespondenceMultilevel1975, gell-mannClassicalEquationsQuantum1993}.  In this context, several attempts have been made to describe quantum systems in purely classical terms, in equilibrium with a bath of quantum fluctuations, described by stochastic noise, which could also encompass {\it e.g.} experimental limitations for measurements of initial conditions. The most famous example is Nelson's stochastic mechanics \cite{nelsonDerivationSchrOdinger1966}. Today, we know that, in the general case, an exact equivalence between quantum mechanics and stochastic classical mechanics is more subtle. However, in the field of open quantum systems, depending on how the quantum properties of a system become apparent when studying it using external probes and how this system is coupled to bath degrees of freedom, approximate solutions of the quantum system can be produced that can provide insight \cite{elyasiResourcesNonlinearCavity2020}. 
It is for this reason that we believe that, at the scale of atomistic spin dynamics simulations, where one can consistently describe thermal fluctuations in a stochastic fashion, it should be possible to approximate the system of spins, thus also describing quantum fluctuations, using a common stochastic approach, though the baths are distinct. In fact, as already mentioned, we have realised this construction for a single quantum spin in a constant magnetic field, in previous work \cite{nussleNumericalSimulationsSpin2023}; taking into account a uniaxial anisotropy represented by an additional quadratic term in the Hamiltonian was done in ref. \onlinecite{nusslePathIntegralSpin}. Essentially, we are performing a systematic expansion in powers of Planck's constant and in powers of temperature and retaining only the leading term in inverse powers of the spin and matching between the partition function on the one hand and the stochastic equation on the other.

The outline of the paper is as follows. In Section \ref{section1}, we begin by recalling results from the exact diagonalisation for a system of two quantum spins $s=1/2$ coupled through an exchange interaction, in the presence of an external field and in equilibrium with a thermal bath. These results are already well known but are essential both to provide a specific reference in our conventions for the expectation values, in order to compare to our atomistic model, as well as as a guideline for computing the overlap between the spin coherent states and the diagonalised spin basis in Section \ref{section3c}. The general method, which in theory is applicable for any principal quantum spin number $s$ as well as for systems of more than 2 spins (for example, chains, 2D lattice, 3D lattice) is briefly described in Appendix~\ref{app:generalsmodel}). Using these results, we produce {\it exact} thermal expectation values for a system of two spins, with $s=1/2$, as a reference
to compare to the results in Section~\ref{section3}. This approach, however, doesn't scale up in a useful way as the Hilbert space size grows quickly with $s$ and number of spin sites $N$, as $\dim=(2s+1)^N$. Moreover, solving the eigenvalue problem becomes even less practical when the spectrum becomes degenerate (which is the case for a chain of spins with $s=1$ already). In Section~\ref{section2},  
we therefore present a procedure for obtaining the expression of an effective classical field, that can be used within the framework of stochastic atomistic spin dynamics simulations based on our earlier work~\cite{nussleNumericalSimulationsSpin2023, nusslePathIntegralSpin}--but being able to handle a more complex Hamiltonian for the initial quantum system due to the interaction between sites. We have named this method Path Integral Spin Dynamics (PISD), as it is inspired from path integral methods in molecular dynamics and here applied in the context of atomistic spin dynamics simulations. This procedure does ``scale up'' in a useful way, as it does not require diagonalising the full $(2s+1)^{2N}$ Hamiltonian operator as a prerequisite. It is an approximation, but provides an expression for the effective field where the spin number $s$ only appears as a parameter rather than an altogether different expression.
In Section~\ref{section3} we compute approximate thermal expectation values, using the path integral atomistic spin dynamics model of Section~\ref{section2}, which we compare to special cases that are exactly solvable, using the results of exact diagonalisation from Section~\ref{section1}.

%%%%%%%%%%%%%%%%%%%%%%%%%%%%%%%%%%%%%%%%%%%%%%%%%%%%%%%%%%%%%%%%%%%%%%%%%%%%%%%%
\section{Exact results for two spins\label{section1}}
%%%%%%%%%%%%%%%%%%%%%%%%%%%%%%%%%%%%%%%%%%%%%%%%%%%%%%%%%%%%%%%%%%%%%%%%%%%%%%%%

The simplest, nontrivial, example involves two spins, $\hat{\bm{S}}^{(1)}$ ($\ket{s^{(1)},m^{(1)}}$, where $m^{(1)}\in \llbracket-s^{(1)},s^{(1)}\rrbracket$) and $\hat{\bm{S}}^{(2)}$ ($\ket{s^{(2)},m^{(2)}}$, where $m^{(2)}\in \llbracket-s^{(2)};s^{(2)}\rrbracket$), coupled by an isotropic Heisenberg exchange interaction and in an external magnetic field $\bm{B}$ along the $\bm{z}$-direction. The Hamiltonian is
\begin{equation}
    {\cal\hat{H}}=-\frac{J}{\hbar^2}\hat{\bm{S}}^{(1)}\cdot\hat{\bm{S}}^{(2)}-\frac{g\muB}{\hbar} B_z\left(\hat{S}^{(1)}_z+\hat{S}^{(2)}_z\right), \label{quantumHamiltonian}
\end{equation}
with $\muB$ the Bohr magneton and $g$ is the gyromagnetic ratio. The standard procedure for diagonalising this two spin system and obtaining an equivalent single spin system with total spin $\hat{\bm{S}}$ and basis $\ket{S,M}$ can be found in appendix \ref{DiagonalisationDetails}.
Using the diagonalised representation we can compute thermal expectation values (or zero-temperature real-time dynamics, upon analytically continuing to real time, though we shall focus on the case of two baths, with which the spins are in equilibrium) from the partition function
\begin{equation}
    {\cal Z}=\mathrm{Tr}\,e^{-\beta {\cal \hat{H}}}=
    \sum_{S,M}\bra{S,M}e^{-\beta {\cal \hat{H}}}\ket{S,M},
\end{equation}
from which we obtain the expression for the thermal expectation values, for instance, the average $z$-component of spin $\hat{S}_z\equiv\frac{1}{2}\left(\hat{S}^{(1)}_z+\hat{S}^{(2)}_z\right)$,
\begin{equation}
    \langle \hat{S}_z\rangle=\frac{1}{2}\frac{\sum_{S,M}\bra{S,M}\left(\hat{S}^{(1)}_z+\hat{S}^{(2)}_z\right)e^{-\beta {\cal \hat{H}}}\ket{S,M}}{\sum_{S,M}\bra{S,M}e^{-\beta {\cal \hat{H}}}\ket{S,M}}.\label{quantumThermalExpectation}
\end{equation}

Simple exponentiation of the diagonal elements of \eqref{singleSpinDiag} leads to
\begin{equation}
    \langle \hat{S}_z\rangle = \frac{\hbar}{2}\frac{ e^{\beta g\muB B_z}- e^{-\beta g\muB B_z}}{e^{-\beta J}+1+e^{-\beta g\muB B_z}+e^{\beta g\muB B_z}}.
\end{equation}

\begin{figure}
    \centering
    \includegraphics[width=0.5\textwidth]{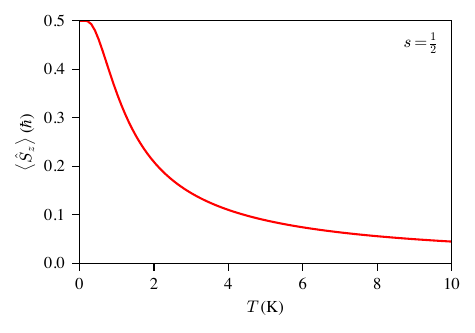}
    \caption{Expectation value $\braket{\hat{S}_z}$ defined in \eqref{quantumThermalExpectation} for two spins with $s=\frac{1}{2}$ as a function of temperature. Exact result for $J = g \muB B_z$, $\bm{B}=(0,0,1)$~T.}\label{exactspinhalf}
\end{figure}

The results for an electron-like particle with $g = 2.00231930436256$, $\beta=1/(k_B T)$, $k_B=1.380649\times10^{-23}~$J~K$^{-1}$ and $\muB = 9.2740100783 \times 10^{-24}$~J~T$^{-1}$ are shown in figure~\ref{exactspinhalf}.
Here we see the decaying ferromagnetic order as the temperature increases. The inflection point is due to the present case of a two-level system (spin $s=1/2$); the thermal spin fluctuations must overcome an initial energy barrier before the alignment with the external field can be destabilised.

Scaling this approach, of rewriting the system in its equivalent single spin incarnation, up to large numbers of spins and larger spin quantisation numbers, encounters two key obstacles.
The first is that this diagonalisation only works for a field that is along a constant direction that we can pick as the quantisation axis, which means this is not the way to proceed for a more general field dependence. The second is that even though the Hamiltonian remains Hermitian regardless of the system size or the value of the spin, this approach quickly becomes impractical. For higher value of the spin, the eigenspectrum displays degeneracies, and the diagonalisation becomes much more difficult, to the point of becoming impractical. Finding the eigenvalues of a matrix of size $N\times N$ amounts to finding the roots of a polynomial of order $N$, which is only exactly/analytically possible for $N\leq4$, unless the polynomial can be factorised. Moreover, having degenerate eigenvalues means that eigenvectors must be hand-picked to ensure that they are orthogonal. So to proceed with our long-term goal of representing thousands of spins, we prefer for every quantum spin to be mapped to a {\it classical} equivalent magnetic moment for our enhanced atomistic spin dynamics model and use these exact results as an exact comparison for small systems only.

%%%%%%%%%%%%%%%%%%%%%%%%%%%%%%%%%%%%%%%%%%%%%%%%%%%%%%%%%%%%%%%%%%%%%%%%%%%%%%%%
\section{Constructing the effective field for PISD simulations\label{section2}}
%%%%%%%%%%%%%%%%%%%%%%%%%%%%%%%%%%%%%%%%%%%%%%%%%%%%%%%%%%%%%%%%%%%%%%%%%%%%%%%%

Our starting point remains the Hamiltonian for two spins $\hat{\bm{S}}^{(1)}$ and $\hat{\bm{S}}^{(2)}$, coupled by isotropic Heisenberg exchange with exchange constant $J$, although now with a magnetic field, $\bm{B}$, in a general direction
\begin{equation}
    \hat{\cal H} = -\frac{J}{\hbar^2}\hat{\bm{S}}^{(1)}\cdot\hat{\bm{S}}^{(2)}-\frac{g\muB}{\hbar}\bm{B}\cdot\left(\hat{\bm{S}}^{(1)}+\hat{\bm{S}}^{(2)}\right).\label{HamiltonianSCS}
\end{equation}
We build the partition function in the coherent spin states basis
\begin{equation}
{\cal Z}=\int \prod_{i=1}^2d\mu(z^{(i)})\bra{z^{(1)}z^{(2)}}e^{-\beta \hat{\cal H}}\ket{z^{(1)}z^{(2)}},\label{PartitionSCS}
\end{equation}
where, assuming that $s^{(1)} = s^{(2)} = s$ (two spin sites can have different values of $s$, and while it doesn't prevent applying the following procedure, it makes the intermediate combinatorial steps more difficult to resolve analytically), the integration measure is given by
\begin{equation}
    d\mu(z^{(i)})=\frac{2s+1}{\pi}\frac{dz^{(i)}}{\left(1+|z^{(i)}|^2\right)^2},
\end{equation}
and the spin coherent states are defined as
\begin{equation}
\begin{aligned}
        &\ket{z^{(1)}z^{(2)}}\equiv\\
        &\frac{\left(1+|z^{(1)}|^2\right)^{-s}}{\left(1+|z^{(2)}|^2\right)^{s}}\sum_{p,p'=0}^{2s}\begin{pmatrix}
            2s\\p
        \end{pmatrix}^{\frac{1}{2}}
        \begin{pmatrix}
            2s\\p'
        \end{pmatrix}^{\frac{1}{2}}z^{(1)p}z^{(2)p'}\ket{p,p'},
\end{aligned}
\end{equation}
with $\{z^{(1)}, z^{(2)}\} \in \mathbb{C}^2$ and where we also defined
\begin{equation}
    \ket{p_1,p_2}\equiv\ket{s_1, s_1-p_1; s_2, s_2-p_2}.
\end{equation}

To recover the classical limit (we emphasize that this is mainly for reference, and the aim of the current work is to go beyond this), one simply neglects all non commuting terms which arise in the Hamiltonian \eqref{HamiltonianSCS}, i.e.
\begin{equation}
    \braket{\hat{S}_x^{(i)l}\hat{S}_y^{(i)m}\hat{S}_z^{(i)n}} \approx \braket{\hat{S}_x^{(i)}}^l \braket{\hat{S}_y^{(i)}}^m \braket{\hat{S}_z^{(i)}}^n,
\end{equation}
where we have defined
\begin{equation}
    \braket{\hat{A}}\equiv\bra{z^{(1)}z^{(2)}}\hat{A}\ket{z^{(1)}z^{(2)}}.
\end{equation}
The reason this is the classical limit is that it does correspond to taking $\hbar s\to 0.$
Recalling \cite{nussleNumericalSimulationsSpin2023}
\begin{equation}
\begin{aligned}
    	\hat{S}^{(i)}_-\ket{p_i,p_j}&=\hbar\sqrt{(2s-p_i)(p_i+1)}\ket{p_i+1,p_j}\\
    	\hat{S}^{(i)}_+\ket{p_i,p_j}&=\hbar\sqrt{p_i(2s-p_i+1)}\ket{p_i-1,p_j}\\
    	\hat{S}^{(i)}_z\ket{p_i,p_j}&=\hbar(s-p_i)\ket{p_i,p_j},
\end{aligned}    
\end{equation}
as well as
\begin{equation}
\begin{aligned}
    	\hat{S}^{(i)}_x&=\frac{\hat{S}^{(i)}_++\hat{S}^{(i)}_-}{2}\\
    	\hat{S}^{(i)}_y&=\frac{\hat{S}^{(i)}_+-\hat{S}^{(i)}_-}{2i}
     \label{sxandsydef},
\end{aligned}
\end{equation}
and then using the stereographic projection which provides a $z^{(i)}\to\bm{n}^{(i)}$ mapping
\begin{equation}
    \left\{
    \begin{aligned}
    n_x^{(i)}&=\frac{z^{(i)}+\bar{z}^{(i)}}{1+|z^{(i)}|^2}\\
    n_y^{(i)}&=-i\frac{z^{(i)}-\bar{z}^{(i)}}{1+|z^{(i)}|^2}\\
    n_z^{(i)}&=\frac{1-|z^{(i)}|^2}{1+|z^{(i)}|^2}
    \end{aligned}
    \right.\label{mappingBlochSphere},
\end{equation}
where $\bm{n}^{(i)}$ are unit vectors, we can map the initial quantum problem in terms of states and operators, to a {\it classical} model of two interacting spin vectors. This yields the partition function
\begin{equation}
		{\cal Z}_\textrm{classical}=\int \prod_{i=1}^2 d\nu(\bm{n}^{(i)})e^{-\beta {\cal H}_\textrm{classical}},
\end{equation}
with the usual classical Heisenberg Hamiltonian
\begin{equation}
    {\cal H}_\textrm{classical} = -J s^2\bm{n}^{(1)}\cdot \bm{n}^{(2)}-g\muB s\bm{B}\cdot\left(\bm{n}^{(1)}+\bm{n}^{(2)}\right),
\end{equation}
and the integration measure restricts the integral to all states on the Bloch sphere for each individual spin
\begin{equation}
	d\nu(\bm{n}^{(i)})=\frac{2s+1}{4\pi}\delta(1-{\bm{n}^{(i)}}^2)d^3n^{(i)},
\end{equation}
where $\delta(x)$ is the Dirac delta function ensuring that the integration is performed on the Bloch sphere, $\bm{n}^2=1$.

To go beyond the classical limit, we obtain an $N$-th order approximate value for the partition function of our system by expanding the operator exponential from the partition function \eqref{PartitionSCS} as a series in the matrix elements (this is by no means a trivial feat and more details on reasonable convergence can be found in Appendix~\ref{app:convergence})
\begin{equation}
\begin{aligned}
	&\bra{z^{(1)}z^{(2)}}e^{-\beta 
    \hat{\cal H}}\ket{z^{(1)}z^{(2)}}\\
    &\approx 1 + \sum_{k=1}^N\frac{(-1)^k\beta^k}{k!} \bra{z^{(1)}z^{(2)}}\hat{\cal H}^k\ket{z^{(1)}z^{(2)}}\\
    &\equiv 1+F[\beta,N].
\end{aligned}\label{seriesExpansion}
\end{equation}
All required matrix elements, up to third order, for computation along with the relevant commutation relations can be found in Appendix~\ref{app:matrix_elements}. Higher orders can be computed symbolically using the python software compendium provided\cite{nussleSourcesPathIntegral2025}.

Now we aim to write the integrand of the partition function \eqref{PartitionSCS} as a unique exponential, that will allow us to identify the effective Hamiltonian. To this end,  we  rewrite \eqref{seriesExpansion} as
\begin{equation}
    1+F[\beta,N] = e^{\ln\left(1+F[\beta,N]\right)}.\label{explntrick}
\end{equation}

At this stage, there are two ways of proceeding. The first is to define the effective Hamiltonian directly from this expression. If the expansion is taken to high enough order, or  there is an exact expression for the matrix elements, i.e. a closed expression for the sum \eqref{seriesExpansion}, for $N=\infty$, as was the case in our previous work for a single spin \cite{nusslePathIntegralSpin}, the effective Hamiltonian is
\begin{equation}
    {\cal H}_\textrm{eff}[N,\{z^{(i)}, \bar{z}^{(i)}\}]\equiv-\frac{1}{\beta}\ln\left(1+F[\beta,N]\right).
    \label{eq:exactHamOrderN}
\end{equation}

The second approach is to compute an expression (for a given order $N$) for the effective Hamiltonian by performing a Taylor series expansion of the exponent of \eqref{explntrick} for $\beta\to0$, which results in a high-temperature approximation.
\begin{equation}
    \ln\left(1+F[\beta,N]\right)\approx \sum_{k=1}^N\frac{(-1)^{k+1}}{k}F[\beta,N]^k,
\end{equation}
which always yields an expression which can be factorised over $\beta$, such that we can define
\begin{equation}
    {\cal H}^{\textrm{high-T}}_\textrm{eff}[N,\{z^{(i)}, \bar{z}^{(i)}\}]\equiv-\frac{1}{\beta}\sum_{k=1}^N\frac{(-1)^{k+1}}{k}F[\beta,N]^k.\label{eq:TaylorHeff}
\end{equation}
A sample computation up to third order is presented in Appendix \ref{app:matrix_elements}. 

Using the mapping \eqref{mappingBlochSphere}, we thus obtain an expression in terms of the two spin coherent state unit vectors $\bm{n}^{(i)}$
\begin{equation}
    {\cal H}_\textrm{eff}[N,\{z^{(i)}, \bar{z}^{(i)}\}]\Rightarrow{\cal H}_\textrm{eff}[N,\{\bm{n}^{(i)}\}].
\end{equation}
From this, we deduce the effective field for use in our atomistic spin dynamics simulation by using the usual expression \cite{skubicMethodAtomisticSpin2008} for the effective field
\begin{equation}
	\bm{B}^{(k)}_{\textrm{eff}}[N]\equiv-\frac{1}{\mu_s}\bm{\nabla}_{\bm{n}^{(k)}}{\cal H}_\textrm{eff}[N,\{\bm{n}^{(i)}\}],\label{effField}
\end{equation}
where $\mu_s=g\muB s$.

We do not provide expressions for the effective Hamiltonians and fields explicitly in this paper as they are much too long to be displayed. They are, however, readily obtainable by computation and printing out, using the python software package in the compendium \cite{nussleSourcesPathIntegral2025}. 

Now that we have constructed the effective Hamiltonian for the atomistic simulations, we shall use it to compute approximate thermal expectation values in the case of an exactly diagonalisable system of two spins, for a constant magnetic field along the quantisation axis, and compare results to the exact quantum results.

%%%%%%%%%%%%%%%%%%%%%%%%%%%%%%%%%%%%%%%%%%%%%%%%%%%%%%%%%%%%%%%%%%%%%%%%%%%%%%%%
\section{Path integral spin dynamics results\label{section3}}
%%%%%%%%%%%%%%%%%%%%%%%%%%%%%%%%%%%%%%%%%%%%%%%%%%%%%%%%%%%%%%%%%%%%%%%%%%%%%%%%

In this section, we begin by recalling essential aspects of atomistic modelling and how to compute thermodynamic averages from the spin dynamics trajectories in section \ref{ASD}, to compare to the expectation values in Section \ref{section1}. We then present results for two distinct cases:
\begin{itemize}
    \item In section \ref{approximateCase} we compute the effective field up to a given order for the exponential series \eqref{seriesExpansion}, the most general case where this series cannot be computed exactly. This is for generic values of $s$ or for a varying field that cannot be taken to coincide with the quantisation axis. Here we also introduce a better method for expanding the exponential series, by expanding around the classical limit of our system. This yields more accurate results than the direct series expansion, at a given approximation order for the exponential series.
    \item In section \ref{section3c} we take advantage of the closedness of the exponential series for fixed values of $s$, when the field direction and quantisation axis coincide, and obtain the effective field exactly by computing the overlap between the diagonalised spin basis and the spin coherent states basis for fixed selected values of $s$. This yields results valid for the whole temperature range.
\end{itemize}

\subsection{Constructing the stochastic atomistic simulation}\label{ASD}

Once an expression for the effective field has been derived (e.g. \eqref{effField} for the high-temperature field from the Taylor expansion method), we proceed in the usual fashion \cite{skubicMethodAtomisticSpin2008} for atomistic spin dynamics simulations by computing dynamical trajectories using the Landau-Lifshitz-Gilbert equation
\begin{equation}
	\dot{\vec{n}}^{(i)}=-\frac{\gamma}{1+\alpha^2}\left(\vec{n}^{(i)}\times\vec{B}^{(i)}_\textrm{eff}+\alpha\vec{n}^{(i)}\times\left(\vec{n}^{(i)}\times\vec{B}^{(i)}_\textrm{eff}\right)\right),
    \label{eq:llg}
\end{equation}
where $\gamma=g\muB/\hbar$ is the gyromagnetic ratio and $\alpha$ is the dimensionless Gilbert damping parameter. We introduce a stochastic field for the thermal fluctuations in the system, $\bm{\eta}$, defined by its first two moments
\begin{equation}
    \begin{aligned}
    \braket{\eta^{(i)}_\mu(t)}&=0\\
    \braket{\eta^{(i)}_\mu(t)\eta^{(j)}_\nu(t')}&=\frac{2\alpha\delta_{ij}\delta_{\mu\nu}\delta(t-t')}{\beta\mu_s\gamma},
    \end{aligned}
\end{equation}
where $\mu$ and $\nu$ stand for the cartesian components of the stochastic field $\bm{\eta}$. The stochastic field is then added to the effective field derived in \eqref{effField}
\begin{equation}
    \bm{B}^{(k)}_{\textrm{eff}}[N]\rightarrow\bm{B}^{(k)}_{\textrm{eff}}[N]+\bm{\eta}.
\end{equation}
We compute thermal averages after an initial relaxation period of $5$~ns, averaging over $N_t$ time samples and $N_s$ independent realisations of the noise and with a system of $N$ spins
\begin{equation}
	\braket{\hat{S}_z} = C \braket{n_z}\equiv  C\frac{1}{N}\frac{1}{N_S}\frac{1}{N_t} \sum_{i=1}^{N}\sum_{j=1}^{N_S} \sum_{k=1}^{N_t} n^{(i)}_{j,z}(t_k).\label{atomisticAverage}
\end{equation}
where $C$ is a normalisation constant to ensure the atomistic results, which are averages from the unit vector $\bm{n}^{(i)}$, are comparable to the spin expectation value which depends on the spin number $s$. For simulations of the classical limit, we simply have $C\hbar s$, for all other approximations, this factor becomes $C=\hbar \left(s+1\right)$. This is related to the classical limit approximating $s(s+1) \approx s^2$. For a more detailed discussion of this issue see the appendix of ref. \onlinecite{nussleNumericalSimulationsSpin2023}.
We integrate the equations of motion numerically using a symplectic algorithm preserving the structure of the phase space\cite{thibaudeauThermostattingAtomicSpin2012}.

At this stage, one can use the code in the compendium \cite{nussleSourcesPathIntegral2025} to produce results for the high-temperature expansion method, as has been done in previous work \cite{nussleNumericalSimulationsSpin2023,nusslePathIntegralSpin}. In practice, however, this approximation scheme tends to not scale very well with increasing exchange strength (see appendix \ref{app:convergence} for a discussion of this issue), especially for low temperatures. There are, however, two simple tricks which can improve this: (a) making an educated guess at what the appropriate classical limit of our quantum system is, and evaluating the effective field as a difference to this limit, and (b) not using the high temperature Taylor series for the effective field, hence yielding an exact expression for the field, to the given order $N$ of the expansion \eqref{eq:exactHamOrderN}.

\subsection{Exact field as difference series from classical limit}\label{approximateCase}

We take two steps to improve the quality of the results and their scaling with both the temperature and the Heisenberg exchange interaction $J$ (or whichever the dominant energetic term is in the Hamiltonian).

\begin{figure}
    \centering
    \includegraphics[width=0.5\textwidth]{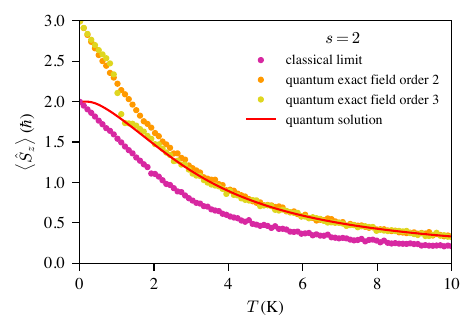}
    \caption{Expectation value $\braket{\hat{S}_z}$ for spin $s=2$ as a function of temperature with $J=g\muB B_z$ and $\alpha = 0.5$, $\bm{B}=(0,0,1)$~T. Atomistic results for classical limit (purple dashed line) and 2nd (orange dashed line) and 3rd (yellow dashed line) order quantum corrections from exact field as difference to classical limit, compared to quantum results (red solid line) from exact diagonalisation method.}\label{spin2exactvsQASDfromdiff}
\end{figure}

Firstly, the Hamiltonian from the Taylor series we take for the expression of the effective Hamiltonian is useful in a pedagogical sense, as it makes the relation to the classical limit more obvious and provides simpler, polynomial expressions \eqref{eq:TaylorHeff} for the effective Hamiltonian/field, but it is not essential. It actually makes results worse. Indeed, when taking the series for the exponential operator, we are simply taking the definition of the exponential of the operator, whilst postulating that this series converges (we emphasize again that this isn't obvious as discussed in Appendix \ref{app:convergence}). This means that taking higher order terms into account enhances convergence towards the correct solution at all temperatures, even if practically, higher temperature are easier to capture as they make the {\it argument} of the exponential smaller. In taking the Taylor series of the logarithm expression however, we explicitly impose that the results are only valid at high temperatures, and in fact there is no requirement to do this, so we will see what happens if we skip this step. This is what we call quantum exact field in figures \ref{spin2exactvsQASDfromdiff}, \ref{fig:exactspinhalfvsexactASD} and \ref{fig:AFcoupling}. In the case where the exponential series is exactly computable as in figures \ref{fig:exactspinhalfvsexactASD} and \ref{fig:AFcoupling}, the field expression is a closed expression and the field is genuinely exact, whereas it is only exact up to the given order in the exponential series for figure \ref{spin2exactvsQASDfromdiff}.

The second step is that we know that, at least for high enough temperatures, the classical limit (when properly normalised) yields qualitatively correct results, so it makes sense to rewrite matrix elements as
\begin{equation}
\begin{aligned}
        \bra{z}e^{-\beta\hat{\cal H}}\ket{z}=e^{-\beta {\cal H}_\textrm{classical}}\bra{z}e^{-\beta\left(\hat{\cal H} -{\cal H}_\textrm{classical}\right)}\ket{z}\\
        \approx e^{-\beta {\cal H}_\textrm{classical}}\bigg(1-\beta \bra{z}\hat{\cal H} -{\cal H}_\textrm{classical}\ket{z}\\
        +\frac{\beta^2}{2}\bra{z}\left(\hat{\cal H} -{\cal H}_\textrm{classical}\right)^2\ket{z}+\dots\bigg),\label{eq:difference_expansion}
\end{aligned}
\end{equation}
and evaluate the effective Hamiltonian from this expression instead. For a discussion on the convergence of this approach, see Appendix~\ref{app:convergence}.

The results for two spins with $s=2$ are given in figure \ref{spin2exactvsQASDfromdiff} with an integration timestep of $5\times10^{-6}$ ns, $N_s=5$ realisations and an average time of $10$ ns (after $5$ ns equilibration time). The order of correction is the number of terms in the Taylor expansion of \eqref{eq:difference_expansion}.
Here we can see that this approach yields promising results. The first correction (orange-dashed curve) already coincides with the quantum solution around $T=4K$ and the second correction (yellow-dashed curve) for $T=1K$.
We will now briefly discuss a method to evaluate these thermal expectation values more accurately by explicitly imposing, as is used for computing the exact quantum expectation values, that the field is constant and chosen to be aligned with the quantisation-axis. This will enable us to derive an exact expression for the effective Hamiltonian, and hence should provide results, valid over the whole temperature range.

\subsection{Effective Hamiltonian for the exactly solvable case }\label{section3c}

In the  case of two spins interacting via an isotropic Heisenberg Hamiltonian with addition of a Zeeman term, under the restriction that the magnetic field and the quantisation axis coincide \eqref{quantumHamiltonian}, we have seen that the Hamiltonian is readily diagonalisable (of course, in principle, it is diagonalisable for any value of spin $s$, despite the procedure becoming more and more cumbersome with increasing dimension of the Hilbert space). This has some more practical use for us namely that we can rewrite the integrand of the partition function \eqref{PartitionSCS}  as
\begin{equation}
\begin{aligned}
        &\bra{z^{(1)}z^{(2)}}e^{-\beta\hat{\cal H}}\ket{z^{(1)}z^{(2)}}\\
        &=\bra{z^{(1)}z^{(2)}}e^{-\beta\hat{\cal H}}\sum_{S,M}\ket{S,M}\braket{S,M|z^{(1)}z^{(2)}}\\
        &=\sum_{S,M}e^{-\beta \lambda_{S,M}}|\braket{S,M|z^{(1)}z^{(2)}}|^2,
\end{aligned}
\end{equation}
where $\lambda_{S,M}$ are the eigenvalues of ${\cal H}$ in the $\ket{S,M}$ basis. Using this, one can obtain an exact expression for the effective Hamiltonian as
\begin{equation}
    {\cal H}_\textrm{eff}=-\frac{1}{\beta}\ln\left(\sum_{S,M}e^{-\beta \lambda_{S,M}}\left|\braket{S,M|z^{(1)}z^{(2)}}\right|^2\right).\label{exactEffField}
\end{equation}

\begin{figure}
    \centering
    \includegraphics[width=0.5\textwidth]{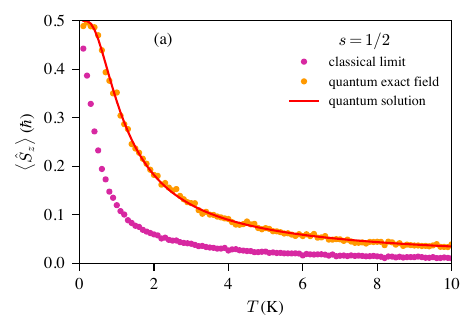}\\
    \hfill\includegraphics[width=0.5\textwidth]{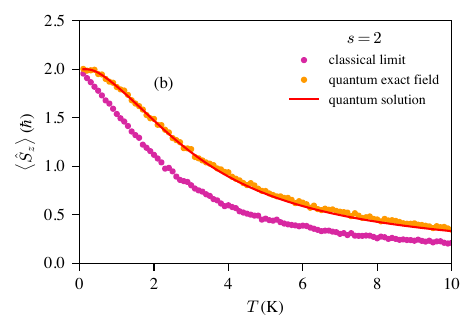}\hfill
    \caption{Expectation value $\braket{\hat{S}_z}$ for two spins with (a) $s=\frac{1}{2}$ and (b) $s=2$, as a function of temperature. Atomistic simulation result (orange dots) from exact effective field \eqref{exactEffField} vs exact result for $J = g \muB B_z$, $\bm{B}=(0,0,1)$~T (solid red line). }\label{fig:exactspinhalfvsexactASD}
\end{figure}

The results are displayed in figure \ref{fig:exactspinhalfvsexactASD} with an integration timestep of $5\times10^{-6}$ ns for sub-figure(a) and $4\times10^{-6}$ ns for sub-figure(b), $N_s=5$ realisations and an average time of $10$ ns (after $5$ ns equilibration time). Here we can indeed see that in this case the whole temperature range is accurately sampled, with the caveat that one requires a constant direction of the magnetic field. In practice, this method is useful for computing most equilibrium thermodynamics quantities such as the Curie temperature, for large spin systems where the applied field can be safely assumed to be constant.

\begin{figure}
    \centering
    \includegraphics[width=0.5\textwidth]{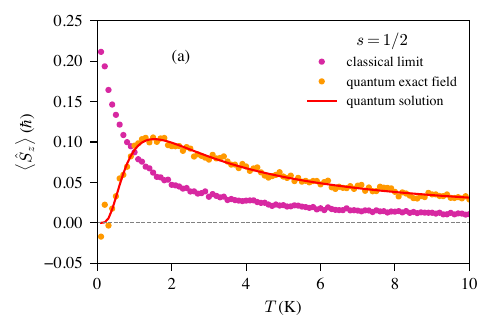}\\
    \hfill\includegraphics[width=0.5\textwidth]{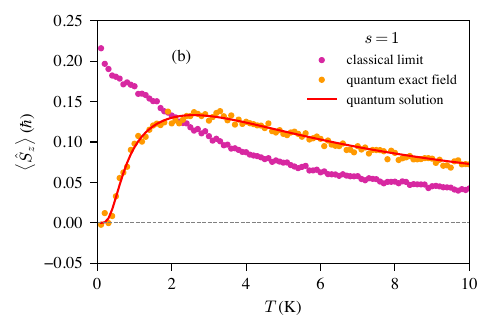}\hfill
    \caption{Expectation value $\braket{\hat{S}_z}$ for two antiferromagnetically coupled spins with (a) $s=\frac{1}{2}$ and (b) $s=1$, as a function of temperature. Atomistic simulation result (orange dots) from exact effective field \eqref{exactEffField} vs exact result for $J = -2g \muB B_z$, $\bm{B}=(0,0,1)$~T (solid red line).}\label{fig:AFcoupling}
\end{figure}

Of particular note is that our method is readily extended to the case of antiferromagnetic coupling with no additional difficulty and provides the correct quantum expectation value as can be seen in figure~\ref{fig:AFcoupling} with an integration timestep of $7\times10^{-7}$ ns, $N_s=5$ realisations for subfigure (a) and $N_s=10$ realisations for subfigure (b)  and an average time of $10$ ns (after $5$ ns equilibration time). Whereas obtaining results for quantum expectation values for antiferromagnetic systems is famously tricky due a family of problems often collectively referred to as ``the sign-problem'', in the context of Quantum Monte Carlo methods for example, where antiferromagnetic systems can lead to negative probability issues\cite{alexandruComplexPathsSign2022}. Figure~\ref{fig:AFcoupling} also gives a clear demonstration of the qualitative difference in the quantum expectation values for antiferromagnets compared to classical models. In the purely classical case (purple-dotted curve, with only thermal fluctuations) the two spins will try to anti-align in the plane normal to the applied field, slightly canting towards the applied field, with decreasing alignment as thermal fluctuations increase, this configuration does not exist for the two quantum spins with $s=1/2$. The ground state for the quantum system is a combination of $\ket{\uparrow\downarrow}$ and $\ket{\downarrow\uparrow}$ states, for which once an initial energy barrier is overcome, increasing thermal fluctuations will be biased to favour spin flips to align with the external field, until the exchange coupling is overcome and then we are back to the thermal decay of the alignment of the spins with the external field. This is perfectly reproduced by the effective classical stochastic model (orange-dotted curve)

%%%%%%%%%%%%%%%%%%%%%%%%%%%%%%%%%%%%%%%%%%%%%%%%%%%%%%%%%%%%%%%%%%%%%%%%%%%%%%%%
\section{Conclusion}
%%%%%%%%%%%%%%%%%%%%%%%%%%%%%%%%%%%%%%%%%%%%%%%%%%%%%%%%%%%%%%%%%%%%%%%%%%%%%%%%

In this work, we expanded the scope of our previous model for simulating quantum spin systems using atomistic spin dynamics and an enhanced effective field which captures thermal and quantum fluctuations. Whereas previously, our model was limited to a single spin interacting with an external field \cite{nussleNumericalSimulationsSpin2023}, or with a uniaxial, quadratic anisotropy term \cite{nusslePathIntegralSpin}, here we have expanded it to include
isotropic Heisenberg exchange and the magnetic field need not coincide with the quantisation axis. This would be relevant for a time-dependent magnetic field for example, though this remains to be spelled out, since the current formalism only applies for computing static expectation values in the presence of quantum and thermal fluctuations at equilibrium. 

As before, we begin by expressing the quantum thermal partition function in the spin coherent states basis so as to provide a continuous, integral definition. At this stage, we need to approximate the matrix elements in the partition function (unless an exact expression is computable). This we do by recalling the series definition of the operator exponential. In essence, this expansion is required to capture non-commutativity due to the curved geometry of the spin phase space and the quantum fluctuations described respectively by the $1/s$ and $\hbar s$ expansions in the introduction. The accuracy of this expansion is depending on the scale of the dominant energetic term in the quantum Hamiltonian being either $\beta g\mu_B\norm{\vec{B}}$ or $\beta J$. We have presented a method where the effective Hamiltonian is computed as a difference to the classical limit which provides a better approximation scheme than the direct expansion from ref. \onlinecite{nussleNumericalSimulationsSpin2023}. We then use the stereographic projection to map the two spins to their corresponding unit {\it spin coherent state vector}. At this stage, we can use the effective field as we would in any other standard atomistic spin dynamics simulation. In the special case where the field is constant and can hence be chosen along the quantisation-axis of the system, we have seen that it is possible to provide a method valid for the whole temperature range. Moreover, we have shown that this also holds for antiferromagnetic systems, for which even the most sucessful methods for computing expectation values of quantum systems, such as quantum Monte Carlo, are often faced with the so-called ``sign-problem''~\cite{alexandruComplexPathsSign2022} While the expressions for the effective fields are impractical to include in this document, they are readily available to be printed out using the software compendium~\cite{nussleSourcesPathIntegral2025}.

At this stage, what needs to be investigated is expanding this method to systems of more than 2 spins. Indeed, whereas this mapping from our initial quantum spin system to a stochastic atomistic spin dynamics simulation can be done exactly in this case, provided that the external field is aligned with the quantisation axis, this is no longer the case for $N$ spin systems, or more precisely, an exact mapping from a quantum system with nearest neighbour interaction only would require all-to-all interactions from the effective model, which would very quickly become impractical. However, insights from open quantum systems and density matrix renormalisation group (DMRG) methods~\cite{hallamTensorNetworkDescriptions2019}, however, seem to indicate that there are approximations that can be consistently made, especially for ``mildly''-entangled systems, as the requirement for these all-to-all interactions to be taken into account is essentially a marker of how entangled a system is. At this stage, we envision as a first approximation to neglect interactions between more than 2 sites and readily generalise the procedure provided in the current work.

This is why in future work we will investigate the effects of our approach for larger atomistic simulations, coupling it to more elaborate thermostatting techniques, which are also built on solid quantum-mechanical grounds\cite{andersQuantumBrownianMotion2022, cerisolaQuantumClassicalCorrespondence2024, berrittaAccountingQuantumEffects2024}, as well as investigating systems that display frustration.
%%%%%%%%%%%%%%%%%%%%%%%%%%%%%%%%%%%%%%%%%%%%%%%%%%%%%%%%%%%%%%%%%%%%%%%%%%%%%%%%

\section*{Acknowledgments}
This work was supported by the Engineering and Physical Sciences Research Council [grant number EP/V037935/1]. JB acknowledges funding from a Royal Society University Research Fellowship. TN acknowledges funding from the Royal Society. JB and TN acknowledge support from a Royal Society International Exchanges 2023 Cost Share (JSPS). TN gratefully acknowledges support from the ``International Network for Spintronics: From Material Development to Novel Energy Efficient Technologies'' funded by EPSRC [grant number EP/V007211/1]. IS acknowledges support from the Royal Society through a summer student scholarship. The authors thank K. Yamamoto for many fruitful discussions and insights.

\section*{Data Access}
Python code and output data to reproduce all results and figures reported in this paper are available from the Zenodo repository: \textit{Sources for: Path integral spin dynamics with exchange and external field} \url{https://doi.org/10.5281/zenodo.14930303}~\cite{nussleSourcesPathIntegral2025}. The repository contains:
\begin{itemize}
    \item Python code for diagonalisation and computing the corresponding thermal expectation values.
    \item Python code to generate analytic expressions derived herein.
    \item Python code to perform enhanced atomistic spin dynamics calculations with the quantum effective fields.
    \item Python scripts to reproduce all the figures.
\end{itemize}
The software and data are available under the terms of the MIT Licence.

\section*{Author Contributions}
% Based on CRediT (Contributor Roles Taxonomy) https://credit.niso.org/
% Example CRediT author statement: https://www.elsevier.com/authors/policies-and-guidelines/credit-author-statement

\textit{Thomas Nussle}: conceptualization, methodology, investigation, software, writing - original draft. \textit{Stam Nicolis}: methodology, writing - review and editing. \textit{Iason Sofos}: exact diagonalisation, software and investigation, writing - review and editing. \textit{Joseph Barker}: conceptualization, methodology, software, data curation, writing - review and editing, funding acquisition.

\appendix

\section{Exact diagonalisation for two-spins}\label{DiagonalisationDetails}

It is possible to recast the two-spin system define by \eqref{quantumHamiltonian}, as an equivalent single-spin system, defined by a state vector $\ket{S,M}$, where $S\in\llbracket|s^{(1)}-s^{(2)}|,s^{(1)}+s^{(2)}\rrbracket$ and $M\in\llbracket-S,S\rrbracket$), using the Clebsch-Gordan coefficients \cite{dirlClebschGordanCoefficients1979}. In terms of the resulting total spin $\hat{\bm{S}}=\hat{\bm{S}}^{(1)}+\hat{\bm{S}}^{(2)}$, the Hamiltonian becomes
\begin{equation}
\begin{aligned}
    \hat{\cal H}&=-\frac{J}{2\hbar^2}\left(\hat{\bm S}^2-s^{(1)}(s^{(1)}+1)-s^{(2)}(s^{(2)}+1)\right)\\
    &-\frac{g \muB B_z}{\hbar}\hat{S}_z.
\end{aligned}
\end{equation}
If we take the example of $s=1/2$, this amounts to a change of basis from $\ket{1/2,\pm 1/2;1/2,\pm 1/2}\equiv\{\ket{\uparrow \uparrow},\ket{\uparrow \downarrow},\ket{\downarrow \uparrow},\ket{\downarrow \downarrow}\}$ represented as
\begin{equation}
\begin{array}{cc}
    \begin{array}{c}
      \bra{\uparrow \uparrow} \hspace{30pt}\bra{\uparrow \downarrow} \hspace{15pt}\bra{\downarrow \uparrow} \hspace{25pt}\bra{\downarrow \downarrow} \\    \end{array}\vspace{3pt}                 &        \\
    \left(\begin{array}{cccc}
\displaystyle -g\muB B_z -\frac{J}{4}    & 0 & 0 &  0 \vspace{2pt} \\
          0    & \displaystyle \frac{J}{4} & \displaystyle -\frac{J}{2} &  0 \vspace{2pt}\\ 
          0    & \displaystyle -\frac{J}{2} & \displaystyle \frac{J}{4} &  0 \vspace{2pt}\\ 
          0    & 0 & 0 & \displaystyle g\muB B_z -\frac{J}{4}\\
    \end{array}\right) & 
        \begin{array}{c}
          \ket{\uparrow \uparrow}\vspace{10pt} \\ \ket{\uparrow \downarrow} \vspace{10pt}\\ \ket{\downarrow \uparrow} \vspace{10pt}\\ \ket{\downarrow \downarrow}
        \end{array}           
\end{array}
\end{equation}
to the new basis $\ket{S,M}\equiv\{\frac{1}{\sqrt{2}}\left(\ket{\uparrow \downarrow}-\ket{\downarrow \uparrow}\right),\ket{\downarrow \downarrow},\frac{1}{\sqrt{2}}\left(\ket{\uparrow \downarrow}+\ket{\downarrow \uparrow}\right),\ket{\uparrow \uparrow}\}\equiv\{\ket{0,0},\ket{1,-1} \ket{1,0} \ket{1,1}\}$, which is represented by
\begin{equation}
\begin{array}{cc}
\hspace{-20pt}\setlength{\arraycolsep}{10pt}
    \begin{array}{cccc}
      \bra{0 ,0} & \bra{1 ,-1} & \bra{1 ,0} & \bra{1 ,1}
    \end{array}\vspace{3pt}                 &        \\
    \left(\begin{array}{cccc}
          \displaystyle \frac{3J}{4}    & 0 & 0 &  0 \\
          0    & \displaystyle g\muB B_z - \frac{J}{4} & 0 &  0 \\ 
          0    & 0 & \displaystyle -\frac{J}{4} &  0 \\ 
          0    & 0 & 0 & \displaystyle -g\muB B_z -\frac{J}{4}\\
    \end{array}\right) & 
        \begin{array}{c}
          \ket{0 ,0}\vspace{10pt} \\ \ket{1,-1} \vspace{10pt}\\ \ket{1,0} \vspace{10pt}\\ \ket{1,1}.
        \end{array}           
\end{array}\label{singleSpinDiag}
\end{equation}

As this is a diagonal matrix, the quantum problem has been solved.

\section{Generalisation for any spin s and for larger systems}\label{app:generalsmodel}
For general values of $s$, one first has to compute the components of the spin operators given by
\begin{equation}
    \left\{
        \begin{matrix}
            (S_x)_{jl} & \displaystyle = \frac{[s(s+1)-j(j-1)]^{\frac{1}{2}}\delta_{jl+1}}{4s}\\
            & \displaystyle +\frac{[s(s+1)-j(j+1)]^{\frac{1}{2}}\delta_{jl-1}}{4s}\\
            \displaystyle(S_y)_{jl} & \displaystyle = \frac{[s(s+1)-j(j-1)]^{\frac{1}{2}}\delta_{jl+1}}{4is}\\
            & \displaystyle -\frac{[s(s+1)-j(j+1)]^{\frac{1}{2}}\delta_{jl-1}}{4is}\\
            \displaystyle(S_z)_{jl} & \displaystyle = j\delta_{jl}
        \end{matrix}
    \right.,
\end{equation}
where $\{j,l\}\in \{-s,-s+1,\dots,s-1,s\}^2$.
From these, one then constructs the Hamiltonian as
\begin{equation}
\begin{aligned}
            &{\cal\hat{H}}=-\frac{g\muB}{\hbar} B_z\bigoplus_{i}\hat{S}^{(i)}_z\\
            &-\frac{J}{\hbar^2}\sum_{\braket{ij}}\left(\hat{S}_x^{(i)}\bigotimes\hat{S}_x^{(j)}+\hat{S}_y^{(i)}\bigotimes\hat{S}_y^{(j)}+\hat{S}_z^{(i)}\bigotimes\hat{S}_z^{(j)}\right).
\end{aligned}
\label{quantumHamiltoniangeneral}
\end{equation}
where $\braket{ij}$ stands for a sum over the nearest neighbour $j$ for every spin site $i$, $\bigoplus$ is the direct product, and $\bigotimes$ is the Kronecker product. The Hamiltonian \eqref{quantumHamiltoniangeneral} constructed in such a fashion is always real and symmetric hence, in theory, always diagonalisable, even in the case of a chain or array of spins of principal number $s>1/2$. In practice, however, finding such a diagonalised matrix analytically quickly becomes intractable, especially when the eigen spectrum becomes degenerate. However, using numerical diagonalisation of matrices, and with sufficient computational resources, one can achieve exact results for larger Hilbert spaces, especially when taking into account symmetry arguments \cite{weiExactDiagonalization16site2023, linExactDiagonalizationQuantumspin1990, schnackExactDiagonalizationTechniques2023}. A python software package which provides exact diagonalisation results (not numerical diagonalisation) for two spins for arbitrary spin $s$ \cite{nussleSourcesPathIntegral2025} is made available alongside the atomistic spin dynamics code, mainly to serve as a reference. In practice, this code has been tested for $s \leq 3$ and higher values may take considerable time to solve, or never provide results. For larger spin, we highly recommend numerical diagonalisation instead, as well as a more thorough symmetry analysis.

\section{Matrix elements and commutation relations}\label{app:matrix_elements}
We present here, for reference, the matrix elements of the spin operators, in the basis of coherent states, highlighting the double expansion in powers of $\hbar s\equiv{\sf s}$ and $1/s$ and the fundamental property that, to any fixed order, $L_\mathrm{max},$ in ${\sf s}$ the moments 
$\langle S_+^{k_1}S_-^{k_2}S_z^{k_3}\rangle,$ with $k_1+k_2+k_3=l\leq L_\mathrm{max},$ are polynomials in $1/s$ of order $l-1$. We focus on the expressions for the moments, $\langle S_+^{k_1}S_-^{k_2}S_z^{k_3}\rangle,$ with $k_1+k_2+k_3=l\leq 3,$ that were used in our calculations.

The matrix elements of $\hat{S}_z^{(i)}$ to $O({\sf s})$ are given by
\begin{equation}
	% &\braket{\hat{S}^{(i)}_+}=2\hbar s\frac{z^{(i)}}{1+|z^{(i)}|^2}\\
	% &\braket{\hat{S}^{(i)}_-}=2\hbar s\frac{\bar{z}^{(i)}}{1+|z^{(i)}|^2}\\
	\braket{\hat{S}^{(i)}_z}=\hbar s\frac{1-|z^{(i)}|^2}{1+|z^{(i)}|^2}={\sf s}\frac{1-|z^{(i)}|^2}{1+|z^{(i)}|^2}
\end{equation}
We note that it doesn't receive a correction to first order in  $1/s,$ consistent with the property that the dependence on $1/s$ for this moment is a constant; the non-trivial property is that the moment can be consistently  normalized, so that this constant is equal to 1.

\begin{widetext}
The matrix elements to $O({\sf s}^2)$ are given by
\begin{align}
\braket{\hat{S}_z^{(i)2}}={\sf s}^2\left\{\left(\frac{1-|z^{(i)}|^2}{1+|z^{(i)}|^2}\right)^2+\frac{2}{s}\frac{|z^{(i)}|^2}{(1+|z^{(i)}|^2)^2} \right\}\\
	 \braket{\hat{S}^{(i)2}_+}=%\hbar^2 2s(2s-1)\frac{z^{(i)2}}{(1+|z^{(i)}|^2)^2}\\
  {\sf s}^2\left(4-\frac{2}{s}\right)\frac{z^{(i)2}}{(1+|z^{(i)}|^2)^2}\\
	 \braket{\hat{S}^{(i)2}_-}=%\hbar^2 2s(2s-1)\frac{\bar{z}^{(i)2}}{(1+|z^{(i)}|^2)^2}\\
  {\sf s}^2\left(4-\frac{2}{s}\right)\frac{\bar{z}^{(i)2}}{(1+|z^{(i)}|^2)^2}\\
% \braket{\hat{S}^{(i)2}_z}=%\hbar^2 s^2 \left(\frac{1-|z^{(i)}|^2}
       % {1+|z^{(i)}|^2}\right)^2+2s\hbar^2\frac{|z^{(i)}|^2}%{(1+|z^{(i)}|^2)^2}\\   
    \braket{\hat{S}^{(i)}_+\hat{S}^{(i)}_-}=%\frac{2 s \hbar^2\left(2 s |z^{(i)}|^2+1\right)}{\left(| z^{(i)}| ^2+1\right)^2}\\
    {\sf s}^2\left(2|z^{(i)}|^2+\frac{1}{s}\right)\frac{2}{(1+|z^{(i)}|^2)^2}\\
    \braket{\hat{S}^{(i)}_-\hat{S}^{(i)}_z}=%\frac{2 s\hbar^2 \bar{z}^{(i)} \left(|z^{(i)}|^2 (1-s)+s\right)}{\left(| z^{(i)}| ^2+1\right)^2}\\
    {\sf s}^2\left(1-|z^{(i)}|^2+\frac{1}{s}|z^{(i)}|^2\right)\frac{2\bar{z}^{(i)}}{(1+|z^{(i)}|^2)^2}\\
    \braket{\hat{S}^{(i)}_+\hat{S}^{(i)}_z}=%\frac{2 s \hbar^2z^{(i)} %\left(-1+s-s | z^{(i)}| ^2\right)}{\left(| z^{(i)}| ^2+1\right)^2}
    {\sf s}^2\left(1-|z^{(i)}|^2-\frac{1}{s}\right)\frac{2z^{(i)}}{(1+|z^{(i)}|^2)^2}
\end{align}

The matrix elements to $O({\sf s}^3)$ are given by
\begin{align}
    &\braket{\hat{S}^{(i)3}_z}=-\hbar^3\frac{s \left(|z^{(i)}| ^2-1\right) \left(s^2 \left(|z^{(i)}| ^4+1\right)-2 ((s-3) s+1) |z^{(i)}|^2\right)}{\left(|z^{(i)}| ^2+1\right)^3}\\
    \\
	&\braket{\hat{S}^{(i)}_+\hat{S}^{(i)}_-\hat{S}^{(i)}_z}=\hbar^3\frac{2 s \left(-2 (s-1) s |z^{(i)}| ^4+(s (2 s-3)+2) |z^{(i)}|^2+s\right)}{\left(|z^{(i)}| ^2+1\right)^3}\\
 \\
    &\braket{\hat{S}^{(i)}_-\hat{S}^{(i)}_-\hat{S}^{(i)}_z}=-\hbar^3\frac{2 s (2 s-1) \bar{z}^{(i)2} \left((s-2) |z^{(i)}|^2-s\right)}{\left(|z^{(i)}| ^2+1\right)^3}\\
    \\
    &\braket{\hat{S}^{(i)}_+\hat{S}^{(i)}_+\hat{S}^{(i)}_z}=-\hbar^3\frac{2 s (2 s-1) z^{(i)2} \left(s |z^{(i)}|^2-s+2\right)}{\left(|z^{(i)}| ^2+1\right)^3}\\
    \\
    &\braket{\hat{S}^{(i)}_+\hat{S}^{(i)}_z\hat{S}^{(i)}_z}=\hbar^3\frac{2 s z^{(i)} \left(s^2 |z^{(i)}| ^4+(-2 (s-2) s-1) |z^{(i)}|^2+(s-1)^2\right)}{\left(|z^{(i)}| ^2+1\right)^3}\\
    \\
    &\braket{\hat{S}^{(i)}_-\hat{S}^{(i)}_z\hat{S}^{(i)}_z}=\hbar^3\frac{2 s \bar{z}^{(i)} \left((s-1)^2 |z^{(i)}| ^4+(-2 (s-2) s-1) |z^{(i)}|^2+s^2\right)}{\left(|z^{(i)}| ^2+1\right)^3}\\
    \\
    &\braket{\hat{S}^{(i)}_+\hat{S}^{(i)}_+\hat{S}^{(i)}_-}=\hbar^3\frac{4 s (2 s-1) z^{(i)} \left(s |z^{(i)}|^2+1\right)}{\left(|z^{(i)}| ^2+1\right)^3}\\
    \\
    &\braket{\hat{S}^{(i)}_+\hat{S}^{(i)}_-\hat{S}^{(i)}_-}=\hbar^3\frac{4 s (2 s-1) \bar{z}^{(i)} \left(s |z^{(i)}|^2+1\right)}{\left(|z^{(i)}| ^2+1\right)^3}
\end{align}
\end{widetext}
and the transcription as ${\sf s}^3P^{(2)}(1/s)$ isn't as illuminating, though it is obvious from these expressions.

For the operators $\hat{S}^{(i)}_\pm$ it is possible to show that there exists a  closed expression for their $N-$th order moments, for any $N,$ given by
\begin{align}
	\braket{\hat{S}^{(i)N}_+}=\hbar^N\frac{(2s)!}{(2s-N)!}\frac{z^{(i)N}}{(1+|z^{(i)}|^2)^N}\\
	\braket{\hat{S}^{(i)N}_-}=\hbar^N\frac{(2s)!}{(2s-N)!}\frac{\bar{z}^{(i)N}}{(1+|z^{(i)}|^2)^N}
\end{align}
These can, also, be recast in the form $(\hbar s)^N P_\pm^{(N-1)}(1/s).$

By using the commutation relations
\begin{equation}
\left\{
\begin{matrix}
        [\hat{S}^{(i)}_z,\hat{S}^{(i)}_\pm]=\pm\hbar\hat{S}^{(i)}_\pm\\
        [\hat{S}^{(i)}_+,\hat{S}^{(i)}_-]=2\hbar\hat{S}^{(i)}_z
\end{matrix}
\right.,
\end{equation}
the expressions for these moments, as well as the definition \eqref{sxandsydef}, one can compute the approximation for the effective Hamiltonian up to third order; in fact, the provided Python package does just this and can be used to print an expression in \LaTeX~for all relevant matrix elements in terms of $\hat{S}^{(i)}_x$, $\hat{S}^{(i)}_y$ and $\hat{S}^{(i)}_z$ in terms of the components of $\bm{n}^{(i)}$, as well as the effective Hamiltonian and corresponding field. By construction, the code can also be used to generate these for any quadratic Hamiltonian up to fifth order. However it must be kept in mind that going up to order higher than three requires significant memory for the symbolic derivation of the effective Hamiltonian, essential to the rest of this procedure. The more complex the initial Hamiltonian, the more memory- and time-consuming this procedure will be.

\section{Convergence analysis}\label{app:convergence}

Expanding an exponential operator as a series is not trivial, as one needs to ensure that this series definition converges, and moreover that only a few terms are sufficient to approximate it, if we wish to use this series as an efficient way of simulating our quantum system, by sampling a {\it classical} Hamiltonian. A safe way to verify this \cite{hiaiTraceNormConvergence1995, suzukiConvergenceGeneralDecompositions1994, barbuNonlinearSemigroupsDifferential1976, rudinFunctionalAnalysis1991}, is to check that
\begin{equation}
    \beta\norm{\hat{\cal H}}_{\infty}\ll 1
\end{equation}
where $\norm{\hat{\cal H}}_{\infty}$ is the supremum norm which in our case, as the Hamiltonian is diagonalisable in the case of the field being constant and chosen along the quantisation axis, can be understood as the largest eigenvalue for the operator $\hat{\cal H}$. If we take the example of the exactly diagonalised result from \eqref{singleSpinDiag} then, if we take for example $J=g\muB B_z$ we require that
\begin{equation}
    \left|\frac{3J}{4k_B}\right|=\left|\frac{3g\muB B_z}{4k_B}\right|\ll T
\end{equation}
specifically in our case this means that this method can safely be expected to be convergent for temperatures such that $T \gg 1.01 K$.

The issue with this approach is that the convergence will  depend quite sensitively  on the value of the exchange constant $J,$  which for realistic materials can be much larger than the applied field. 

\begin{figure}
    \centering
    \includegraphics[width=0.5\textwidth]{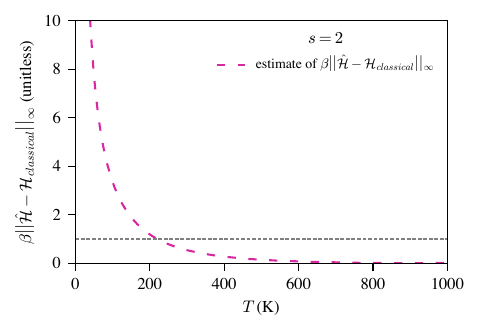}
    \caption{Estimate of the convergence criteria when taking the first term in the exponential expansion \eqref{eq:difference_expansion} for the evaluation of the effective Hamiltonian as a difference to the classical limit of the quantum Hamiltonian, for spin $s=2$ as a function of temperature, with $J=100g\muB$ and $\alpha = 0.5$. The grey dashed line indicates $y=1$ below which this approach converges.}\label{fig:difClassicalLimit}
\end{figure}

A more reasonable approach is to take as appropriate measure the norm of the difference between  the effective Hamiltonian  and  its classical limit so that, despite the influence of the relevant constants in the quantum Hamiltonian, these will also appear in the corresponding classical limit, hence one can hope for a larger domain of convergence and fewer orders of corrections required for a good degree of approximation. At this point a good estimate for the convergence of the exponential series will be
\begin{equation}
    \beta\norm{\left(\hat{\cal H}-{\cal H}_{\textrm{classical}}\right)}_{\infty}\ll 1.
\end{equation}
As a rough estimate, we use the difference between the highest eigenvalue of the quantum Hamiltonian and the corresponding value of the classical Hamiltonian at a given temperature. Results are given in figure \ref{fig:difClassicalLimit} for a much higher value of $J=100g\muB B_z$ with $B_z = 1T$ which is the right order of magnitude for standard ferromagnets. 

In practice this means that the first correction of this method should be sufficient to provide a reasonable approximation starting from around $T=200K$. We would like to emphasize here that a single quantum spin (or rather a few, if we include concepts such as entanglement) is much less likely to behave {\it classically} than a collection of them. Indeed a larger system provides more routes for the quantum degrees of freedom to interact with through fluctuation and dissipation when in contact with a bath, and therefore can be expected to be more readily approximated by a classical model.

\sloppy
\bibliography{PISD.bib}

\end{document}